\documentclass[conference,a4paper,10pt]{IEEEtran}
\usepackage{amsmath,amssymb}
\usepackage{graphicx}
\usepackage{cite}

\usepackage{amsmath, amssymb, amsfonts}
\usepackage{bm}
\usepackage{cite}
\usepackage{hyperref}
\usepackage{mathtools}
\usepackage{xcolor}
\usepackage{balance}
\usepackage{caption} 
\usepackage{microtype}

\newcommand{\MI}{\ensuremath{\mathcal{I}}}
\newcommand{\E}{\ensuremath{\mathcal{H}}}

\begin{document}

\title{Model-Aware Rate--Distortion Limits for Task--Oriented Source Coding}

\author{
    \IEEEauthorblockN{Andriy Enttsel, Vincent Corlay}
    \IEEEauthorblockA{
        Mitsubishi Electric R\&D Centre Europe, Rennes, France \\
        \{a.enttsel, v.corlay\}@fr.merce.mee.com
    }
}

\maketitle

\begin{abstract}
Task-Oriented Source Coding (TOSC) has emerged as a paradigm for efficient visual data communication in machine-centric inference systems, where bitrate, latency, and task performance must be jointly optimized under resource constraints. While recent works have proposed rate--distortion bounds for coding for machines, these results often rely on strong assumptions on task identifiability and neglect the impact of deployed task models. In this work, we revisit the fundamental limits of single-TOSC through the lens of indirect rate--distortion theory. We highlight the conditions under which existing rate--distortion bounds are achievable and show their limitations in realistic settings. We then introduce task model-aware rate--distortion bounds that account for task model suboptimality and architectural constraints. Experiments on standard classification benchmarks confirm that current learned TOSC schemes operate far from these limits, highlighting transmitter-side complexity as a key bottleneck.
\end{abstract}
\begin{IEEEkeywords}
task-oriented source coding, coding for machines, indirect rate--distortion theory, classification
\end{IEEEkeywords}
\section{Introduction}
\label{sec:intro}

Edge devices, mobile phones, and autonomous vehicles generate enormous amounts of visual data, including images, videos, and point clouds. In many applications, the visual content needs to be processed to extract semantically meaningful information. Considering the scale of the data, the processing is nowadays automated and performed by vision models operating under the machine analysis paradigm. 

To enable this, data are often compressed and dispatched to the cloud to exploit its higher computational capabilities.
Conventional source coding aims to preserve visual (perceptual) fidelity of the original content. In contrast, task-oriented source coding (TOSC), also known as coding for machines (CfM), aims to compress data efficiently without compromising the accuracy of the subsequent tasks, such as classification \cite{Singh2020ICIP, Matsubara2022WACV}, segmentation \cite{Ahuja2023CVPR}, and object detection \cite{Yuan2022MIPR}. 

A related framework is the collaborative human-machine setting, where the codec jointly optimizes task performance and visual quality \cite{Duan2020TIP, Yang2021TM, Yang2024TPAMI}. Settings targeting multiple tasks have also been studied  \cite{Alvar2021TIP, Matsubara2025WACV}.

Two fundamental quantities that characterize compression performance are the rate accounting for the amount of the conveyed information, which is typically measured in visual codecs in bits per pixel (bpp), and the fidelity of the compression, usually accounted for by a proper task-dependent distortion measure. Rate and distortion are inherently competing objectives, and their trade-off was formalized in information theory by the rate--distortion function \cite{Berger1971}.

\textbf{Related work}:  In this work, we further elaborate on the rate--distortion analysis of the single-task-oriented source coding recently addressed in \cite{Harell2025TPAMI, Bajic2025MIRP}. 
A common paradigm in neural-network-based TOSC systems is to consider a model designed for the specific task (task model) and to split it between the transmitter and the receiver, thereby defining a task front end and a task back end. In this setting, the transmitted signal is the intermediate features produced by the front end.
While the splitting point is a fundamental design choice, the authors in \cite{Harell2025TPAMI} proved that, from an information-theoretic perspective, this choice does not affect the optimal rate–distortion performance, which instead depends only on the task and the corresponding distortion metric.
Nevertheless, in \cite{Bajic2025MIRP} it has been shown that current state-of-the-art (SOTA) single-task TOSC methods operate orders of magnitude away from the theoretical limit in terms of the bitrate needed to achieve a certain level of accuracy. The main cause for this was tied to the notion of the removal of irrelevant information formalized in information theory by the information-bottleneck method \cite{Tishby1999AACCCC}. 

\textbf{Contribution}: The bound proposed in \cite{Bajic2025MIRP} was derived from the standard rate--distortion function addressing the source coding of the task-relevant signal. Based on the work of \cite{Liu2021ISIT} on semantic communication rate--distortion, and in particular on the indirect rate--distortion function \cite{Dobrushin1962TIT, Witsenhausen1980TIT}, 
\emph{i}) we highlight the implicit assumptions and scenarios that may limit the significance of the bound in \cite{Bajic2025MIRP};
\emph{ii}) we propose three new bounds that are not only task-dependent but also explicitly account for the task model;
\emph{iii}) by comparing state-of-the-art TOSC schemes to the proposed bounds, we identify the main bottleneck to achieving superior performance as the computational and architectural complexity constraints at the transmitter.

\section{Rate--distortion limits of single-task TOSC}
\label{sec:ird}

In this section, we characterize the fundamental limits of single-task TOSC through the lens of \emph{indirect rate--distortion} (iRD) theory \cite{Dobrushin1962TIT, Witsenhausen1980TIT}. We motivate iRD as the correct information-theoretic limit for TOSC in contrast to other commonly adopted rate–distortion formulations. We then clarify how standard coding strategies relate to iRD, and under which assumptions they can be optimal.

Given an observation modality, e.g., an image or a video, the objective of CfM and TOSC is to convey to a remote agent a task-relevant latent variable, e.g., class label, object identity, using the minimum possible communication rate, subject to a constraint on task fidelity.

The signal flow is described by the Markov chain
\begin{equation}
\label{eq:remote_markov}
Y \rightarrow X \rightarrow \hat{Y},
\end{equation}
where $Y$ denotes the unobserved task-relevant variable, $X$ the observation available at the transmitter, and $\hat{Y}$ the (possibly distorted) reconstruction of $Y$ made available at the receiver. 
The channel $Y \rightarrow X$ is imposed by the data-generating process and described by $p(x|y)$. The channel $X \rightarrow \hat{Y}$, described by $p(\hat{y}|x)$, is determined by the TOSC design. Under \eqref{eq:remote_markov}, the optimal rate--distortion trade-off is given by \cite{Dobrushin1962TIT, Witsenhausen1980TIT}
\begin{equation}
\label{op: remote_rd}
\tag{iRD}
R_X(D_Y)
=
\min_{p(\hat{y}|x)} \MI(X;\hat{Y})
\quad \text{s.t.} \quad
\mathbb{E}[d(Y,\hat{Y})] \le D_Y,
\end{equation}
where the mutual information $\MI(X;\hat{Y})$ models the communication rate and $d(y,\hat{y})$ is a task-dependent distortion measure. Problem~\eqref{op: remote_rd} is known as the \emph{indirect (or remote) rate--distortion function} and represents the true fundamental limit of single-task TOSC.

\subsection{Operational bounds via factorized coding strategies}

Directly solving~\eqref{op: remote_rd} is generally intractable in practice, since the observation $X$ representing the visual content is typically high-dimensional and complex. Nevertheless, useful operational bounds can be obtained by restricting the design to factorized coding strategies. For this, we consider the more general processing chain where $X \rightarrow \hat{Y}$ is factorized through an intermediate representation $Z$
\begin{equation}
\label{eq:processing_markov}
Y \rightarrow X \rightarrow Z \rightarrow \hat{Y}.
\end{equation}
Two canonical TOSC strategies arise depending on whether the additional processing is performed at the transmitter or at the receiver.

\subsubsection{Compress-and-estimate}

In the compress-and-estimate  (C\&E) coding scheme, the transmitter first compresses the observation $X$, producing $Z=\hat{X}$, and the receiver subsequently estimates the task variable as $\hat{Y}=g(\hat{X})$  (the estimator $g$ is usually called the task head). The compression phase is described by the classical (direct) rate--distortion function \cite{Cover_2006}
\begin{equation}
	\label{op: direct_rd}
	\tag{RD}
	R_X(D_X)
	=
	\min_{p(\hat{x}|x)} \MI(X;\hat{X})
	\quad \text{s.t.} \quad
	\mathbb{E}[d(X,\hat{X})] \le D_X .
\end{equation}

The task distortion is then computed as $D_Y=\mathbb{E}[d(Y,g(\hat{X}))]$. Since C\&E conveys the full observation $X$, including information that may be irrelevant for the task, the resulting rate--distortion performance constitutes an \emph{achievable upper bound} on the indirect rate--distortion function in \eqref{op: remote_rd}. The C\&E strategy is optimal only in the degenerate case $Y=X$, in which~\eqref{op: direct_rd} reduces to a special case of \eqref{op: remote_rd}. Nevertheless, C\&E can be relevant when the task is unknown at the time of compression and is decided only at the receiver \cite{Kipnis2021TIT}, as pursued by the JPEG-AI standard \cite{Ascenso2023Multim, Alshina2024Multim, Alkhateeb2025TM}.

\subsubsection{Estimate-and-compress}

In the estimate-and-compress (E\&C) strategy, the transmitter first estimates the task variable, $Z=\tilde{Y}=f(X)$ (the estimator $f$ is usually called the task model), and then compresses the estimate. The compression stage is governed by the conventional rate--distortion function for the estimated variable
 \begin{equation}
	\tag{eRD}
	 \label{op: direct_rd2}
	R_{\tilde{Y}}(D_{\tilde{Y}})
	=
	\min_{p(\hat{y}|\tilde{y})} \MI(\tilde{Y};\hat{Y})
	\quad \text{s.t.} \quad
	\mathbb{E}[d(\tilde{Y},\hat{Y})] \le D_{\tilde{Y}} .
\end{equation}
and the relevant task distortion is $D_Y=\mathbb{E}[d(Y,\hat{Y})]$. By compressing a task-related representation instead of the raw pixel domain observation, E\&C may achieve a lower rate for a given task distortion than C\&E, provided that the task estimation is sufficiently accurate. Some notable cases where E\&C is optimal are when $X$ and $Y$ are jointly Gaussian and the distortion $d(y,\hat{y})$ is the mean squared error (MSE)\footnote{In this case $f$ is the minimum mean squared error estimator.} \cite{Liu2021ISIT}, and when the compression is lossless.

\subsection{Identifiability and oracle rate--distortion function}

Assume that the task variable is perfectly \emph{identifiable} at the transmitter, i.e., there exists a deterministic mapping such that $Y=f(X)$, or equivalently, the conditional entropy $\E(Y|X)=0$. In this case, the Markov chain reduces to $X \rightarrow Y \rightarrow \hat{Y}$, and~\eqref{op: remote_rd} simplifies to
\begin{equation}
\label{op: semantic_rd}
\tag{oRD}
R_Y(D_Y)
=
\min_{p(\hat{y}|y)} \MI(Y;\hat{Y})
\quad \text{s.t.} \quad
\mathbb{E}[d(Y,\hat{Y})] \le D_Y .
\end{equation}
This is the conventional rate--distortion function of the task-relevant variable, which we call the oracle rate--distortion function (oRD) and it represents the \emph{absolute lower bound} for lossy compression of $Y$.

To summarize, the following inequality holds:
\begin{equation}
R_Y(D_Y)
\le
R_X(D_Y)
\le
\min\!\left\{
R_X^{\mathrm{C\&E}}(D_Y),\;
R_{\tilde Y}^{\mathrm{E\&C}}(D_Y)
\right\},
\end{equation}
where $R_X(D_Y)$ is defined by \eqref{op: remote_rd}, and the right-hand terms denote the rate--task distortion performance of the C\&E and E\&C strategies.

Although the identifiability condition $\E(Y|X)=0$ is often implicitly assumed in the recent literature \cite{Harell2025TPAMI, Bajic2025MIRP}, it is a strong requirement and must be verified for the task at hand. While this assumption may be approximately satisfied in certain curated benchmarks, it is frequently violated in realistic visual inference scenarios involving noise, ambiguity, or inherent uncertainty\footnote{We discuss this aspect more thoroughly in the Appendix.}.
When it does not hold, the bound~\eqref{op: semantic_rd} is generally \emph{unachievable} (even with a perfect E\&C strategy involving an optimal estimator and an optimal compressor) and does not characterize the true TOSC limit. 

\begin{figure}[t]
        \centering
        \includegraphics[width=\linewidth]{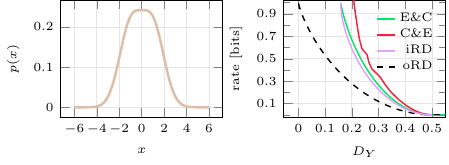}
	\caption{Gaussian mixture model with overlapping classes: probability density (left) and rate–distortion performance (right).}
    \label{fig: GMM}
\end{figure}

\subsection{Motivating example}

To illustrate that the \eqref{op: semantic_rd} bound valid under perfect identifiability can be unattainable and strictly loose w.r.t. \eqref{op: remote_rd}, and that neither C\&E nor E\&C is generally optimal for TOSC, we consider the following example of \eqref{eq:remote_markov}. 
$Y$ is generated as $Y \sim \mathrm{Bernoulli}(q)$ with parameter $q=\mathbb{P}(Y=1)=0.5$ and $X = 2Y - 1 + Z $ with $Z \sim \mathcal{N}(0,1)$. The resulting observation $X$ represents a shifted, Gaussian-noise-corrupted version of the task-relevant variable $Y$ and has marginal density $p(x) = 0.5 p(x|y=0) + 0.5 p(x|y=1)  =  0.5 \mathcal{N}(x; -1, 1) + 0.5 \mathcal{N}(x; 1, 1)$, i.e., a two-component Gaussian mixture model with equiprobable modes shown in Fig.\ref{fig: GMM}~(left).

Since the two modes overlap, no classifier can perfectly discriminate\footnote{In this case, the optimal classifier defined by the maximum a posteriori decision rule evaluates the sign: $f(X) = \operatorname{sign}(X).$} between them and identify $Y$. In Fig.\ref{fig: GMM}~(right), we compare the rate--distortion performance of different TOSC strategies. E\&C and \eqref{op: remote_rd} for this source are derived as in \cite{Liu2021ISIT}. We additionally report the \eqref{op: semantic_rd} specialized for the binary source and the C\&E strategy, which solves \eqref{op: direct_rd} and applies the maximum a posteriori (MAP) estimator on the reconstruction $\hat x$ (further details can be found in the Appendix). 
The figure shows that the hypothesis of perfect identifiability yields a loose and unachievable lower bound on the optimal TOSC performance and highlights the suboptimality of both E\&C and C\&E.

\section{Task model-dependent bounds}

The bounds on TOSC analyzed in \cite{Bajic2025MIRP} are given by the oracle rate--distortion function \eqref{op: semantic_rd}. As discussed in the previous section, this bound relies on the assumption that the task-relevant variable is perfectly identifiable at the transmitter.  

Moreover, \eqref{op: semantic_rd} does not account for practical limitations, such as the use of a specific, potentially suboptimal, task model. When such a task model is deployed, the achievable rate--distortion trade-off is fundamentally affected by the task model estimation error
\begin{equation}
	D_{\text{TM}}=\mathbb{E}\Big[ d\big( Y, \tilde{Y} \big) \Big],
\end{equation}
introduced prior to compression.

In this work, we instead propose bounds that directly target the more realistic indirect rate--distortion function \eqref{op: remote_rd} and explicitly account for a given task model.

Two of these bounds are based on the E\&C strategy.
If the target distortion equals to the irreducible distortion floor, i,e., $D_Y=D_{\text{TM}}$, the optimal strategy within the E\&C framework is to estimate $Y$ and then losslessly compress the estimate, since any lossy compression of the estimate would necessarily introduce additional distortion and result in $D_Y>D_{\text{TM}}$. For larger target distortions $D_Y>D_{\text{TM}}$, let us specialize the Markov chain in \eqref{eq:processing_markov} to the E\&C case:
\begin{equation}
\label{eq: e and c markov}
	Y \rightarrow X \rightarrow \tilde{Y} \rightarrow \hat{Y},
\end{equation}
Under this model, we can immediately improve over $R_{\tilde Y}^{\mathrm{E\&C}}(D_Y)$ by interpreting the output of the estimator $\tilde{Y}$ as the updated observation of the remote source $Y$, obtained via the mapping $X\rightarrow \tilde{Y}$. This leads to the appropriate indirect rate-distortion function:
\begin{equation}
\tag{iE\&C}
\label{op: indirect_rd2}
	R_{\tilde Y}(D_Y)= \min_{p(\hat{y}\mid \tilde{y})} \MI (\tilde{Y}; \hat{Y} )
	\quad \text{s.t.}  \quad  \mathbb{E}[d(Y,\hat{Y})]\le D_Y.
\end{equation}
The latter formulation suggests that, given the source $\tilde{Y}$, improving over E\&C requires the compressor to reconstruct not the estimate $\tilde{Y}$ itself, but rather the task-relevant variable $Y$. We refer to this approach as indirect estimate-and-compress (iE\&C): estimation and compression remain separated, but compression is optimized directly with respect to task distortion. In practice, iE\&C leads to the unsurprising fact that the compressor must be trained with respect to task labels to be more effective\footnote{Knowledge of $Y$ is assumed for training but not during inference.}.

Note that, while iE\&C provides a strictly tighter benchmark than classical E\&C for design spaces constrained by a fixed task model, it remains suboptimal with respect to the unrestricted  \eqref{op: remote_rd} bound:
\begin{equation}
R_Y(D_Y)
\le
R_X(D_Y)
\le
R_{\tilde Y}(D_Y)
\le
R_{\tilde Y}^{\mathrm{E\&C}}(D_Y).
\end{equation}

Finally, the main advantage of iE\&C and E\&C relative to \eqref{op: remote_rd} is that they are operational, as they avoid direct optimization over the high-dimensional source $X$.

\begin{figure*}[t]
\centering
\begin{minipage}[b]{0.255\textwidth}
  \centering
  \includegraphics[width=\linewidth]{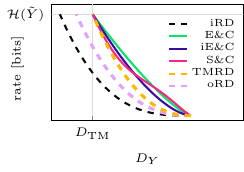}
  \captionof{figure}{Qualitative behavior of the rate--distortion bounds.}
  \label{fig:qualitative}
\end{minipage}\hfill
\begin{minipage}[b]{0.72\textwidth}
\centering
  \includegraphics[width=\linewidth]{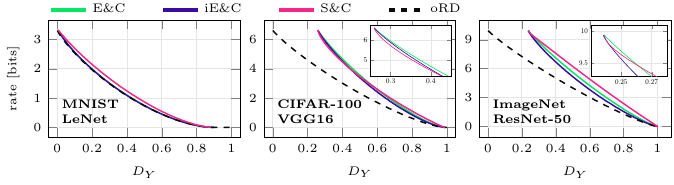}
   \captionof{figure}{Rate--distortion performance across standard classification benchmarks based on common vision task models.}
 \label{fig:rd-bounds}
\end{minipage}

\end{figure*}

\subsection{Classification}
We now focus on the classification task, for which the proposed bounds can be numerically derived using the Blahut--Arimoto (BA) algorithm \cite{Blahut1972TIT, Cover_2006}. In its standard form, the BA algorithm computes the direct rate-distortion function of a discrete memoryless source by solving the Lagrangian relaxation $R_Z + \lambda D_Z$, where $Z \sim p(z)$ is a discrete scalar random variable and $\lambda > 0$ is the Lagrangian multiplier controlling the trade-off.

For classification, the task-relevant feature $Y$ is a discrete random variable on the alphabet $\mathcal{Y}=\{1,\ldots,|\mathcal{Y}|\}$, with prior probabilities $p(y)=\mathbb{P}(Y=y)$, $y\in\mathcal{Y}$ implicitly defined by the task data generation process. The output of the estimator (classification task model) $\tilde{Y}$ takes values on the same alphabet. 

The goal of training the task model is to learn a mapping from $X$ to $\tilde{Y}$ that induces a conditional distribution $p(\tilde{y}\mid y)=\mathbb{P}(\tilde{Y}=\tilde{y}\mid Y=y)$, with $y,\tilde{y}\in\mathcal{Y}$, such that the distortion $D_{\text{TM}}=\mathbb{E}[d(Y, \tilde{Y})]=\mathbb{P}(\tilde{Y}\neq Y)$, with $ d(\tilde{y}, y)=\mathbf{1}\{\tilde{y} \neq y\}$, is minimized.

Given a trained softmax-based classification task model, let $f_{\text{TM}}(x) \in \mathbb{R}^{|\mathcal{Y}|}$ represent the vector of logits at the output of the pre-softmax layer. The model defines class probabilities
\begin{equation}
\label{eq: posterior class}
p(y\mid x)=\mathbb{P}(Y=y \mid X=x)=\operatorname{softmax}\!\left[f_{\text{TM}}(x)\right]_y,
\end{equation}
and the corresponding estimate $\tilde{Y}$ is obtained via the MAP decision rule.
The resulting classifier can be completely characterized by its confusion matrix, which (if normalized per row) represents the empirical estimate of the channel $p(\tilde{y}\mid y)$. For instance, the distortion it induces and the  marginal probabilities can be computed as follows
\begin{align}
\label{eq: dist_class}
	D_{\text{TM}} &= \mathbb{P}(\tilde{Y}\neq Y)
	=\sum_{y \in \mathcal{Y}}\mathbb{P}(\tilde{Y}\neq y\mid Y=y)\,p(y)\\
	&=\sum_{y \in \mathcal{Y}}\left[1-\mathbb{P}(\tilde{Y}=y\mid Y=y)\right]p(y)\\
\label{eq: marginal_class}
	p(\tilde{y})&=\mathbb{P}(\tilde{Y}=\tilde{y})
	=\sum_{y \in \mathcal{Y}} p(\tilde{y},y)
	=\sum_{y \in \mathcal{Y}} p(\tilde{y}\mid y)\,p(y).
\end{align}

\subsubsection{E\&C}
In the case of the E\&C strategy, \eqref{op: direct_rd2} can be numerically solved by the BA algorithm, which takes as input $p(\tilde{y})$ with $\tilde{y}\in \mathcal{Y}$ and, based on $\lambda$ and the distortion function $d(\tilde{y},\hat{y})=\mathbf{1}\{\tilde{y} \neq \hat{y}\}$, numerically computes the rate, distortion and the channel
\begin{equation}
	p(\hat{y}\mid \tilde{y}) = \mathbb{P}(\hat{Y}=\hat{y}\mid \tilde{Y}=\tilde{y}),
\end{equation}
where $\hat{Y}$ is a discrete random variable representing the reconstruction of the classifier output $\tilde{Y}$.
The distortion $\mathbb{E}[d(\tilde{y},\hat{y})]$ computed by BA reflects optimal lossy compression via $p(\hat{y}\mid\tilde{y})$, whereas evaluating $R_{\tilde Y}^{\mathrm{E\&C}}(D_Y)$ requires the task distortion $D_Y$, given by
\begin{align}
    \label{eq: task-dist}
    D_Y &= \mathbb{E}\!\left[d(Y,\hat{Y})\right] = \mathbb{P}(\hat {Y}\neq  Y)\\
    &=\sum_{y \in \mathcal{Y}}\left[1-\mathbb{P}(\hat{Y}=y\mid Y=y)\right]p(y).
\end{align}
Computing \eqref{eq: task-dist} requires evaluating the overall channel $p(\hat{y}\mid y)$, capturing the combined effect of estimation and compression. From the Markov chain in \eqref{eq: e and c markov} $p(\hat{y},\tilde{y},y)=p(\hat{y}\mid \tilde{y})\,p(\tilde{y}\mid y)\,p(y)$ and marginalizing out $\tilde{Y}$, it follows that 
\begin{equation}
    p(\hat{y}\mid y)=\sum_{\tilde y \in \mathcal{Y}} p(\hat{y}\mid \tilde{y})\,p(\tilde{y}\mid y).
\end{equation}

\subsubsection{iE\&C}

To solve \eqref{op: indirect_rd2} via the BA algorithm, the task distortion $d(y,\hat{y})$ must be expressed in terms of the variables $\tilde{Y}$ and $\hat{Y}$, which are the source and the reconstruction seen by the compressor.
This can be achieved by the conditional expectation `trick' \cite{Witsenhausen1980TIT}: it is sufficient to define an effective distortion
\begin{equation}
\label{eq: effective distortion}
	\hat{d}(\tilde{y},\hat{y})=\mathbb{E}\!\left[d(Y,\hat{y}) \mid \tilde{Y}=\tilde{y}\right],
\end{equation}
and by the law of total expectation, this guaranties
\begin{equation}
	\mathbb{E}\!\left[\hat{d}(\tilde{Y},\hat{Y})\right]
	=\mathbb{E}\!\left[d(Y,\hat{Y})\right]
	= D_Y.
\end{equation}
That is, by taking the distortion in \eqref{eq: effective distortion} as input, the BA algorithm minimizes the expected effective distortion and hence equivalently minimizes the original task distortion.
In our case, by the definition of conditional expectation \eqref{eq: effective distortion} can be expressed in closed form 
 \begin{align}
 	\hat{d}(\tilde{y},\hat{y})&=\sum_{y \in \mathcal{Y}}  d(y,\hat{y})\,p(y\mid \tilde{y})=\sum_{ y \in \mathcal{Y}} \mathbf{1}\{y\neq \hat{y}\}\,p(y\mid \tilde{y})\\
 	&=1-\mathbb{P}(Y=\hat{y}\mid \tilde{Y}=\tilde{y}),
 \end{align}
where the posterior 
\begin{equation}
	p(y\mid \tilde{y})
	= \mathbb{P}(Y=y \mid \tilde{Y}=\tilde{y})
	= \frac{p(\tilde{y}\mid y) p(y)}{p(\tilde{y})}.
\end{equation}
 
\subsubsection{Sample \& communicate}

While iE\&C improves over E\&C, it is not a generally optimal strategy for TOSC. For $D_Y > D_{\mathrm{TM}}$, it has been shown in~\cite{Liu2021ISIT} that optimal classification-oriented coding does not rely on hard estimation, but instead exploits soft, uncertainty-aware representations of the task variable. This is captured by the posterior $p(y \mid x)$ in~\eqref{eq: posterior class}, which, for a fixed task model, represents the Bayesian sufficient statistic of $X$ for reconstructing $Y$.
Hence, optimal task-model-constrained TOSC requires solving an indirect rate--distortion problem (we label as TMRD) with respect to 
\begin{align}
    V = \big(p(1 \mid X), \dots, p(|\mathcal{Y}| \mid X)\big) = \operatorname{softmax}[f_{\text{TM}}(X)]
\end{align}
rather than a hard estimate $\tilde{Y}$:
\begin{equation}
\tag{TMRD}
\label{op: tmrd}
R_{V}(D_Y)= \min_{p(\hat{y}\mid v)} \MI(V; \hat{Y})
\quad \text{s.t.} \quad \mathbb{E}[d(Y,\hat{Y})]\le D_Y .
\end{equation}
Unfortunately, $V$ is continuous-valued and high-dimensional, which limits the applicability of the BA algorithm. We therefore leverage this insight to define a new TOSC bound that captures the benefit of uncertainty-aware representations without relying on BA.

Rather than forming a hard decision $\tilde{Y}$ through argmax operation on \eqref{eq: posterior class} and then compressing it, we consider a scenario where one \textit{samples} the posterior and then losslessly \textit{communicates} the sample to the receiver. With this, the reconstruction can be defined as
\begin{equation}
	(\hat{Y}\mid X=x) \sim \operatorname{Cat}\big(\operatorname{softmax}[\lambda f_{\text{TM}}(x)]\big),
\end{equation}
where $\operatorname{Cat}$ denotes a categorical distribution parameterized by the softmax output. The inverse-temperature parameter $\lambda$ is introduced to heuristically control the rate--distortion trade-off.

Operationally, the sample and communicate strategy (S\&C) is an instance of channel simulation \cite{LiFTCIT2024}, or relative entropy coding \cite{Flamich2020NIPS}. Classical results \cite{Bennett_TIT2002} on channel simulation show that the fundamental limit on the required communication rate is given by $\MI(X;\hat{Y})$.

Since S\&C strategy operates at the rate $\MI(X; \hat{Y})$, it directly targets the indirect rate-distortion function defined by \eqref{op: remote_rd} and $\lambda$ allows to control the trade-off. Qualitatively, when $\lambda \rightarrow +\infty$, the softmax operation behaves as argmax, and we recover the lossless operating point. When $\lambda \rightarrow 0$, the sampling operation generates $\hat{Y}$ independently of $X$, minimizing the mutual information between them. Quantitatively, we can analyze the rate-distortion performance by empirically estimating $\MI(X; \hat{Y})$ and $\mathbb{E}[d(Y,\hat{Y})]$ from a given dataset $\mathcal{D}=\{(x_i,y_i)\mid x_i\in\mathcal{X},y_i\in\mathcal{L}\}$ (details are in the Appendix).

In Fig.~\ref{fig:qualitative} we summarize the rate--distortion behaviors by plotting qualitative trends. We highlight the intractable \eqref{op: tmrd} representing the best rate--distortion performance attainable for a fixed task model, and the intractable \eqref{op: remote_rd}, defining the fundamental limit of TOSC. The bound \eqref{op: tmrd} is expected to lie between the absolute lower bound \eqref{op: semantic_rd} and the infimum of the task-model-dependent bounds considered here, while \eqref{op: remote_rd} lies between \eqref{op: semantic_rd} and \eqref{op: tmrd}.

\subsection{Generalizability}
Although the computation of the bounds is presented for canonical classification, this does not necessarily restrict the applicability of the framework. As shown in \cite{Bajic2025MIRP}, tasks such as object detection can be cast as classification over the classes defined by the various configurations of object locations and object classes in an image, allowing the same rate--distortion analysis to apply in principle. The resulting limitations are computational rather than conceptual. In this respect, the advantage of S\&C is that it scales linearly with the number of classes, while BA scales quadratically.
Nevertheless, due to the convexity of \eqref{op: remote_rd}, when the number of classes makes other approaches computationally infeasible one can adopt a simple affine upper bound\footnote{It is asymptotically achievable by a time-sharing (TS) scheme \cite[Ch. 8]{Yeung_2008} where the transmitter with probability $\lambda$ sends the entropy coded estimate and with probability $1-\lambda$ sends nothing. In the latter case, the receiver via MAP sets $\hat y = \arg \max_{y}p(y)$. With the scheduling protocol known to both ends, the rate is $R = \lambda \E(\tilde Y)$ and the distortion is $D_Y = \lambda  D_{\text{TM}} + (1-\lambda ) D_0 $.
} connecting the zero-rate point $(0, D_0) $, where $D_0 = 1-\max_{y}p(y)$, to the lossless-estimation point $(\E( \tilde Y), D_{\text{TM}})$:
\begin{equation}
	\label{eq: naive}
	R^{\text{TS}}_{\tilde Y}(D_Y) =\frac{D_0 - D_Y}{D_0-  D_{\text{TM}}} \E(\tilde Y)  \quad \text{for} \quad D_{\text{TM}}  \leq D_Y  \leq D_0.
\end{equation}

\section{Numerical results}
\subsection{Rate-distortion}
To illustrate the outlined rate-distortion bounds, we consider three common classification benchmarks: MNIST \cite{LeCun1998}, CIFAR-100 \cite{Krizhevsky2009}, and ImageNet \cite{Russakovsky_2015IJCV}. For MNIST we design a simple LeNet-inspired \cite{Lecun_1996ProcIEEE} classifier achieving 99.2\% classification accuracy ($D_\text{TM}=8 \cdot 10^{-3}$). For CIFAR-100, we employ a finetuned VGG-16 \cite{Simonyan_2014arxiv} leading to 74\% classification accuracy ($D_\text{TM}=0.26$). For ImageNet we deploy ResNet-50 \cite{He_2016CVPR} achieving 76.1\% classification accuracy ($D_\text{TM}=0.239$) . In Fig.~\ref{fig:rd-bounds}, for each dataset, we plot E\&C, iE\&C, and S\&C-based bounds. For reference, we also plot the rate-distortion curve corresponding to \eqref{op: semantic_rd}. Since all class labels are uniformly distributed, \eqref{op: semantic_rd} has a closed-form solution provided in the Appendix and corresponds to the bound analyzed in \cite{Bajic2025MIRP}. 
The near-perfect discrimination in the MNIST case (Fig.~\ref{fig:rd-bounds} (left)) implies that the E\&C and iE\&C strategies nearly achieve the absolute lower bound given by \eqref{op: semantic_rd}. Being E\&C nearly optimal makes any other strategy suboptimal, including S\&C. For CIFAR-100 in Fig.~\ref{fig:rd-bounds} (middle), iE\&C always outperforms E\&C, while S\&C achieves better than iE\&C performance for $0.26 \leq D_Y \leq 0.4$ and becomes worse than E\&C for $D_Y \geq 0.57$. 
Finally, for ImageNet illustrated in Fig.~\ref{fig:rd-bounds} (right),  iE\&C provides a substantial improvement over E\&C, while S\&C is highly suboptimal, except in a narrow region around $D_Y \simeq 0.24$. 

We emphasize that, while S\&C incorporates estimation uncertainty via posterior sampling, the rate--distortion trade-off is controlled solely by softmax tempering, which lacks an explicit optimality guarantee and may explain why S\&C becomes suboptimal for $D \gg  D_{\text{TM}}$. Nevertheless, an important empirical result is that S\&C outperforms iE\&C in certain scenarios, thereby confirming the latter’s general suboptimality and indicating potential room for improvement. 

Finally, in Fig.~\ref{fig:sota}, by plotting accuracy--bit per pixel (bpp)\footnote{$\text{bpp} = \text{average bit rate per observation} / \text{\# of pixels per observation}$} curves, we showcase that the current SOTA ResNet-50-based TOSC scheme for classification \cite{Harell2025TPAMI} is not only highly suboptimal w.r.t. to the unachievable bound \eqref{op: semantic_rd}, as already observed in \cite{Bajic2025MIRP}, but also to the proposed task model-dependent bounds. The two SOTA curves correspond to two different model splitting points considered in the rate-accuracy analysis of \cite{Harell2025TPAMI}.

\subsection{Complexity}

While the quality of the estimator limits accuracy, we argue that the gap in rate is primarily due to complexity constraints, which we quantify in terms of model size in bits\footnote{Model size in bits is both a coarse proxy for Kolmogorov complexity \cite{Li2019Springer}, a principled information-theoretic notion of the complexity of a digital object, and is widely adopted in model deployment benchmarks \cite{Apple2024ResNet50}.}. 

The proposed bounds assume that a full ResNet-50 model of size 45.06 MB\footnote{We analyze the model with float 16 precision as it does not bring any loss of performance w.r.t. the standard float 32 precision.} operates at the transmitter. To comply with transmitter-side complexity constraints, the SOTA1 method deploys only the task front end up to Block~0 of Layer~4\footnote{Layer naming follows the convention of \cite{Harell2025TPAMI}; see Fig.~18 in the corresponding supplementary material for a detailed architectural illustration.}; the rest of the model, including the fully connected layer, is executed at the receiver. According to our analysis, this almost halves the transmitter side model size down to 28 MB. The SOTA2 configuration keeps the layers up to Block~4 of Layer~3, further halving the size to 12.16 MB.
Note that the effective complexity reduction is smaller, as we do not account for the additional model employed to entropy code the features. 

\begin{figure}[t]
    \centering
    \includegraphics[width=\linewidth]{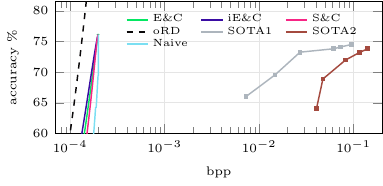}
    \caption{State-of-the-art accuracy versus bits-per-pixel performance, compared with theoretical bounds and a naive E\&C baseline.}
    \label{fig:sota}
\end{figure}

To further support our claim, Fig.~\ref{fig:sota} reports in cyan the performance of a naive yet operational one-shot TOSC baseline. 
Specifically, a reduced codebook is constructed by merging the classes with the $k$ smallest marginal probabilities $p(\tilde y)$, obtained at the output of the classifier, into a single symbol, which is then reassigned to the class with the largest $p(\tilde y)$. The resulting distribution $p(\hat y)$ is entropy coded, and the parameter $k$ controls the rate--distortion trade-off. Compared to the proposed bounds, the rate–-distortion inefficiency due to suboptimal compression of $\tilde Y$ is negligible when contrasted with the performance gap exhibited by the current SOTA method.

Lastly, we note that techniques orthogonal to model splitting can be used to reduce transmitter-side complexity, including pruning and quantization \cite{Gao2019ICML, Cheng2024TPAMI}\footnote{See Table~4 in the supplementary material of \cite{Cheng2024TPAMI} for the supervised and unsupervised pruning techniques applied to ResNet-50.}, with only marginal accuracy losses.

\subsection{Discussion}
The complexity considerations align with prior work on task-aware compression under resource constraints. For instance, in \cite{Matsubara2022WACV, Matsubara2025WACV} the authors empirically show that the achievable rate--accuracy trade-off depends critically on the complexity budget of the edge model. However, those works adopt a systems and learning perspective, whereas our goal is to connect this practical limitation to the fundamental iRD framework and to show that current TOSC schemes operate in a complexity-limited regime that is inherently suboptimal from an information-theoretic standpoint.

In \cite{Bajic2025MIRP}, the performance gap was related to inefficient removal of task-irrelevant information and suboptimal representation learning in current systems, using the information bottleneck (IB) principle \cite{Tishby1999AACCCC}. The IB formulation can be interpreted as a special case of iRD with logarithmic loss distortion \cite{Gunduz_2023}. This perspective is also consistent with the present complexity analysis: due to resource constraints, the information bottleneck is effectively inserted too early in the processing pipeline, which limits the attainable rate--distortion operating point and explains the observed gap to the iRD bound.

\section{Conclusion}
In this work, we revisited the fundamental limits of task-oriented source coding through the lens of the indirect rate--distortion theory. By explicitly accounting for task model suboptimality, we derived model-aware rate--distortion bounds that provide operational benchmarks for practical TOSC systems. Our analysis shows that commonly adopted rate–distortion bounds rely on strong identifiability assumptions that may not hold in realistic settings. Numerical results on standard classification benchmarks demonstrate that current state-of-the-art TOSC schemes operate far from these limits, with transmitter-side complexity and the induced need for model splitting emerging as the dominant bottlenecks.

These findings confirm that rate-distortion analysis alone is insufficient to fully assess TOSC performance and that explicit complexity measures must also be considered. While estimate-and-compress strategies are generally suboptimal, they remain viable in the rate-distortion sense and can, in principle, 
meet practical complexity constraints when paired with standard model-reduction techniques.

\section*{Appendix}
\label{sec: appendix}

\subsection*{Discussion on identifiability in standard classification benchmarks}
\label{subsec: discussion}
The oracle rate--distortion bounds commonly adopted in the literature rely on the implicit identifiability assumption
$\E(Y \mid X) = 0$. This assumption is often motivated by the fact that most vision benchmarks are human-labeled, and that a single ``ground-truth'' label is deterministically assigned to each image.

However, human annotation does not imply identifiability in the information-theoretic sense. A more faithful description distinguishes between the latent semantic variable $Y$  and the observed annotation $\tilde{Y}$, produced by a labeling process operating on the observation $X$. This induces the Markov chain
\begin{equation}
Y \rightarrow X \rightarrow \tilde{Y},
\end{equation}
where $\tilde Y$ is generally a noisy estimate of $Y$. Even when the annotation noise is small and operationally neglected, the underlying conditional entropy $\E(Y \mid X)$ does not vanish.

Moreover, identifiability may fail even in the absence of annotation noise due to irreversible transformations in the data acquisition or preprocessing pipeline. For instance, in low-resolution benchmarks such as CIFAR-100, the downsampling process discards fine-grained visual information, so that multiple distinct semantic configurations may give rise to the same observation $X$, implying $\E(Y \mid X) > 0$.
This phenomenon becomes explicit in corrupted versions of standard benchmarks, where noise, blur, and other perturbations are deliberately injected with varying intensity~\cite{Hendrycks_2018ICLR}.

These considerations highlight that the identifiability condition underlying oracle rate--distortion bounds is a strong modeling assumption, which may hold approximately for curated datasets but is generally violated in realistic visual inference scenarios. In such cases, the oracle rate--distortion function does not characterize the operational limits of task-oriented source coding (TOSC), and the indirect rate--distortion formulation provides the appropriate theoretical framework.
\subsection*{Rate--distortion function of a binary source}
 For $Y \sim \mathrm{Bernoulli}(q)$ we have \cite[Th. 10.3.1]{Cover_2006}
 \begin{equation*}
 	R_Y(D_Y)=
 	\begin{cases}
 		h(q)-h(D_Y), &  \text{if } 0 \le D_Y \le \min(q,1-q),\\
 		0, &  \text{if } D_Y \ge \min(q,1-q),
 	\end{cases}
 \end{equation*}
 where $h(u) = -u \log_2 u - (1 - u)\log_2(1 - u)$ is the entropy of the binary source $Z \sim \mathrm{Bernoulli}(u)$.

\subsection*{C\&E plot details}
To plot the C\&E curve in Fig.~\ref{fig: GMM} (right), we first discretize the source and then apply the Blahut--Arimoto \cite{Blahut1972TIT} algorithm to compute the channel following the rate-distortion function: $
p(\hat{x}\mid x)= \mathbb{P}(\hat{X}=\hat{x}\mid X=x)$.
Considering a reconstructed sample, the optimal classification rule based on the maximum a posteriori estimator gives 
$\hat{y} = \arg \max_{y}\, p(y\mid \hat{x})$
and
$\mathbb{P}(\hat{Y}=y \mid \hat X=\hat{x})=\max_{y}\, p(y\mid \hat{x})$.
The task-relevant distortion is then computed as:
\begin{align*}
D_Y &= \mathbb{P}(\hat{Y}\neq Y)= \sum_{\hat{x}} p(\hat{x})\, \mathbb{P}(\hat{Y}\neq Y\mid \hat{x})\\
&= \sum_{\hat{x}} p(\hat{x})\left[1-\max_{y}\, p(y\mid \hat{x})\right].
\end{align*}
where, due to the Markov chain $Y\rightarrow X\rightarrow \hat{X}$, $p(y,x,\hat{x}) = p(\hat{x}\mid x)p(x\mid y)p(y)$ and the posterior is computed as
\[
p(y\mid \hat x)=\frac{p(y,\hat{x})}{p(\hat{x})}
=\frac{\sum_{x} p(y,x,\hat{x})}{p(\hat{x})}
=\frac{\sum_{\hat x} p(\hat{x}\mid x)p(x\mid y)p(y)}{\sum_{\hat x} p(\hat{x}\mid x)\, p(x)}.
\]

\subsection*{Estimation details of S\&C}
\label{subsec: s and c}
The mutual information can be expressed as $
\MI(X; \hat{Y}) =\E(\hat{Y}) - \E(\hat{Y}\mid X) $.
We first estimate 
\[
p(\hat{y}) \simeq \frac{1}{|\mathcal{X}|} \sum_{x_i\in \mathcal{X}} p(\hat{y}\mid x_i)
\]
to compute the entropy of $\hat{Y}$
\[
\E(\hat{Y}) = -\sum_{ \hat y \in \mathcal{\hat Y}}  p(\hat{y}) \log_2p(\hat{y}).
\]
The conditional entropy is estimated as 
\begin{align*}
	\E(\hat{Y}\mid X) &\simeq \frac{1}{| \mathcal{X} |} \sum_{x_i\in\mathcal{X}} \E(\hat{Y}\mid X=x_i)\\
	&= -\frac{1}{| \mathcal{X} |}  \sum_{x_i\in\mathcal{X}} 
	\left\{ \sum_{ \hat y \in \mathcal{\hat Y}} p(\hat{y}\mid x_i)\log_2p(\hat{y}\mid x_i)\right\},
\end{align*}

To compute the distortion, we rely on the conditional expectation $
\mathbb{E}[d(Y,\hat{Y})]=\mathbb{E}[\hat{d}(X,Y)],
$
where 
\begin{align*}
	\hat{d}(x,y)&=\mathbb{E}[d(\hat{Y},y)\mid X=x]
	= \sum_{ \hat y \in \mathcal{\hat Y}}  d(y,\hat{y})\,p(\hat{y}\mid x)\\
	&= \sum_{ \hat y \in \mathcal{\hat Y}}  \mathbf{1}\{y\neq \hat{y}\}\,p(\hat{y}\mid x)
	=1-\mathbb{P}(\hat{Y}=y\mid X=x).
\end{align*}
and the empirical estimate becomes
\[
\mathbb{E}[d(Y,\hat{Y})] \simeq 1-\frac{1}{|\mathcal{D}|}\sum_{i=1}^{|\mathcal{D}|} \mathbb{P}(\hat{Y}=y_i\mid X=x_i).
\]

\subsection*{Classification rate-distortion function}
For the discrete memoryless source of $|\mathcal{Y}|$ uniformly distributed elements $y \in \{ 1, \dots, |\mathcal{Y}|\}$, i.e., $p(y) = 1 / |\mathcal{Y}|$ $\forall y$, the rate-distortion function is
\begin{equation*}
R_Y(D_Y)=
\begin{cases}
\log_2 |\mathcal{Y}| - h(D_Y) - D_Y \log_2\!\left(|\mathcal{Y}| - 1\right),\\
  &\hspace{-4em} \text{if } 0 \le D_Y \le D_0,\\
0, & \hspace{-4em}  \text{if } D_Y \ge D_0.
\end{cases}
\end{equation*}
where $D_0 = 1 - 1 / |\mathcal{Y}|$.

 \balance
{
\footnotesize
\bibliographystyle{IEEEbib}
\bibliography{refs_abrv.bib}
}
\end{document}